\documentclass[floatfix,11pt,nofootinbib]{revtex4}
\usepackage{amssymb,amsmath}
\usepackage{graphicx,float}
\usepackage{indentfirst}

\newcommand{\beq}{\begin{equation}}
\newcommand{\eeq}{\end{equation}}
\newcommand{\bea}{\begin{eqnarray}}
\newcommand{\eea}{\end{eqnarray}}

\def\({\left(}
\def\){\right)}

\input{tcilatex}
\begin{document}

\title{Axion-fermion Coupling and Dyon Charge as Physical Signatures of a
Space-time Inner Symmetry}
\author{Andr\'{e} Martorano Kuerten}
\email{martoranokuerten@hotmail.com}
\affiliation{Tubar\~{a}o, Santa Catarina, Brazil}

\begin{abstract}
In this paper we intend to complement the identification given in Kuerten \&
Fernandes-Silva \cite{amkafs} (Mod. Phys. Lett. A. Vol. 33, No. 16) which
relates the axion to a metric spinor phase by means of Maxwell's theory in
the Infeld-van der Waerden's $\gamma $-formalism. Thus, we obtain two
alternative identifications: the first focuses on Dirac's theory so that
when obtaining an axion-like phase-fermion coupling, we achieve the first
identification, and the last one investigates the phase behavior under
Peccei-Quinn rotations in order to show that the phase changes as an axion
pseudoparticle. With the formal aspects established, we also study the
semiclassical fermion-photon system to demonstrate that the magnetic
monopole current defined in \cite{amkafs} has dyon charge in flat universe
and acquires a Witten effect form when there is a demand for chiral symmetry.
\end{abstract}

\maketitle

\section{Introduction}

Axion theories have provided several advances in theoretical and
experimental physics, such as quantum chromodynamics, condensed matter,
string theory and dark matter studies \cite%
{zhang,bakas,preskill,abbott,fischler,gondolo}. Peccei \& Quinn \cite%
{peccei,quinn} introduced the axion field to solve the strong problem in 
\textrm{CP} symmetry. In the Ref. \cite{amkafs}, the so-called axion
electrodynamics has played an important role to identify the axion with a
metric spinor phase. Introduced by Wilczek \cite{wilczek}, this new
electrodynamics finds basis on the Lagrangian term%
\begin{equation}
\alpha F^{\mu \nu }F_{\mu \nu }^{\star },  \label{1}
\end{equation}%
where $F_{\mu \nu }$ is the Maxwell tensor, $F_{\mu \nu }^{\star }$ is its
Hodge dual and $\alpha $ the axion field. As usual, in trivial topology
cases, we have $\alpha F^{\mu \nu }F_{\mu \nu }^{\star }=0$. The term (\ref%
{1}) modifies Maxwell inhomogeneous equations. Since then, some
generalizations have been elaborated to extend the Wilczek's theory, and
Tiwari establishes a local dual invariant electrodynamics (\textrm{LDIE})
which naturally includes an axion electrodynamics \cite%
{tiwari,tiwari2,tiwari3,tiwari4}. Instead of the original axion
electrodynamics formulation, the \textrm{LDIE} also modifies homogeneous
equations. The non-observation of magnetic monopoles is justified by
demanding that the axionic correction cancels the magnetic monopole term.

In \cite{amkafs} we see the \textrm{LDIE} that satisfies the Maxwell
equations is identical to that found in the $\gamma $-formalism, when
magnetic monopole currents are properly defined. Those magnetic currents got
inspiration from the definition of geometrical sources for Infeld-van der
Waerden electromagnetic fields given in \cite{kuerten}. In Infeld-van der
Waerden's formalisms \cite%
{waerden,infeld2,infeld,corson,bade,cardoso,cardoso2,cardoso3}, the usual
metric spinor $\varepsilon _{AB}$, used in the $\varepsilon $-formalism
admits phase and scale transformations that do not affect the fundamental
metric structure $g_{\mu \nu }$. In fact, taking into account $\varepsilon
_{AB}\mapsto \left\vert \gamma \right\vert e^{\Theta i}\varepsilon _{AB}$
and $\Sigma _{\mu }^{AA^{\prime }}\mapsto \left\vert \gamma \right\vert
^{-1}\Sigma _{\mu }^{AA^{\prime }}$ , the metric tensor transforms as $%
g_{\mu \nu }\mapsto g_{\mu \nu }$, where $\Sigma _{\mu }^{AA^{\prime }}$ is
a connecting object component (also known as Infeld-van der Waerden symbol).
Therefore, the $\gamma $-formalism defines the metric spinor and the
connecting object in the following manner%
\begin{equation}
\gamma _{AB}\doteqdot \left\vert \gamma \right\vert e^{\Theta i}\varepsilon
_{AB}\text{ \ \ and \ \ }\Upsilon _{\mu }^{AA^{\prime }}\doteqdot \left\vert
\gamma \right\vert ^{-1}\Sigma _{\mu }^{AA^{\prime }}.  \label{2}
\end{equation}%
Thus, each $\gamma \varepsilon $-formalism does have base on its metric
spinor: constant in the $\varepsilon $-formalism and depending locally on
the space-time coordinates in the $\gamma $-formalism.

Formally, the Infeld-van der Waerden's formalisms are based on the
homomorphism $\mathfrak{h}:\mathcal{W}(2,\mathbb{C})\rightarrow \mathcal{SO}%
_{\uparrow }^{+}(1,3)$ between the Weyl $\mathcal{W}(2,\mathbb{C})$ and the
orthochronous proper Lorentz group $\mathcal{SO}_{\uparrow }^{+}(1,3)$.
Sometimes classical world\ theories can be rewritten in a $2$-spinor
version, since there is a homomorphism two to one which relates the linear
group of unimodular complex $2\times 2$ matrices $\mathcal{SL}(2,\mathbb{C})$
to the group $\mathcal{SO}_{\uparrow }^{+}(1,3)$. The group $\mathcal{W}(2,%
\mathbb{C})$ is longer than $\mathcal{SL}(2,\mathbb{C})$ and contains a $%
\mathcal{U}(1)$ freedom, which is symbolically stated by $\mathfrak{h}\left(
e^{i\lambda }\mathfrak{g}\right) =\mathfrak{h}\left( \mathfrak{g}\right) \in 
\mathcal{SO}_{\uparrow }^{+}(1,3)$ where $\mathfrak{g}\in \mathcal{W}(2,%
\mathbb{C})$ \cite{afriat}.

Historically, Infeld and van der Waerden considered the Weyl's work \cite%
{weyl} to implement Dirac's theory in General Relativity. Weyl studied the
relationship between the tetrad formalism for curved space-time and the
parameter of the Dirac $4$-spinor phase transformation to conclude that if
the tetrad varies, the parameter varies as well \cite{afriat,weyl}.
Originally, the theory provided a geometrical origin of the electromagnetic
potential, since it would lead to an imaginary part of the spinor connection
trace that would satisfy the Weyl's principle of gauge invariance.
Unfortunately, there was no consolidation of this idea and it should not be
understood on its original form \cite{honorinf}. In this interpretation, as
the scale/phase couples with each type of fermion, the formalism would imply
a relation between electric charge and spin. However, the neutron disabled
this idea, because it has spin but no electric charge. Furthermore, the
interpretation of the imaginary part of the spinor connection trace impaired
some investigations over Maxwell's theory in the $\gamma $-formalism.
Nowadays, the physical significance of the phase is free to be reinterpreted
on a modern perspective.

In \cite{amkafs}, it considered the potential as an external physical entity
such that it establised the identification $\Theta $ $\sim $ $\alpha $.
However, the identification solely was based on the Maxwell's theory.
Therefore, in the present study, we will take other ways that lead to such
identification. Wanting to repeat a similar result with the derived in \cite%
{amkafs}, in here, we will investigate Dirac's theory in the $\gamma $%
-formalism. The axion-fermion coupling is given by the following Lagrangian
term \cite{raffelt}%
\begin{equation}
\Psi ^{\dagger }\gamma ^{0}\gamma ^{\mu }\Psi \partial _{\mu }\alpha ,
\label{afc}
\end{equation}%
with $\Psi $ being the Dirac $4$-spinor, $\Psi ^{\dagger }\gamma ^{0}$ its
adjoint spinor and $\gamma ^{\mu }$ the Dirac matrices. Exclusively in this
paper, we start showing a similar identification through of the couplings of
the form (\ref{afc}) between some fermions and the space-time phase within
the scope of Dirac's theory in $\gamma $-formalism. Posteriorly, we consider
the Peccei-Quinn transformations $\Psi \mapsto $ $e^{i\zeta \gamma _{5}}\Psi 
$ where the axion behaves as $\alpha \mapsto $ $\alpha -2\zeta $ when such
transformations are taken into account. Keeping this in mind, we use the $2$%
-component chiral fermions to implement the Peccei-Quinn group in the
Infeld-van der Waerden $\gamma $-formalism and so to show that the
space-time phase also transforms as an axion pseudoparticle. Besides these
identifications, we also consider the fermion-photon system and demonstrate
that the magnetic monopole source of \cite{amkafs} carries a dyon charge in
Minkowski space-time as also, on a specific case, it has a Witten effect
form where the space-time phase assumes the axion role.

We will also use $\hbar =c=1$ as the metric signature $(+---)$. Round/square
brackets will indicate the index symmetry/antisymmetry. The paper's
organization will be as it follows. In the section $2$, we will review the
axion-electrodynamics provided by the $\gamma $-formalism. In the section $3$%
, we must obtain the axion-fermion couplings from $\gamma $-formalism and
establish the Weyl-Peccei Quinn transformations to identify the phase with
the axion again. In the section $4$, we will use the Maxwell-Dirac system to
explicitly obtain a magnetic monopole $2$-spinor form and derive its
effective magnetic charge. In addition, we will show that such charge has a
dyon structure and that it assumes a quantized Witten form when requiring a
chiral invariance.

\section{Identifying Axion with Metric Spinor Phase: The Photon Case}

In the present section, we will review the results obtained in \cite{amkafs}%
. The Lagrangian that provides the axion electrodynamics is given by%
\begin{equation}
2F^{\mu \nu }F_{\mu \nu }+4\alpha F^{\mu \nu }F_{\mu \nu }^{\star }+A^{\mu
}j_{\mu },  \label{4}
\end{equation}%
where $\mathcal{L}_{\text{M}}=2F^{\mu \nu }F_{\mu \nu }$ is the usual
Maxwell Lagrangian, $\mathcal{L}_{\text{CS}}=4\alpha F^{\mu \nu }F_{\mu \nu
}^{\star }$ is the Chern-Simons term and $\mathcal{L}_{\text{I}}=A^{\mu
}j_{\mu }$ an interaction term between the gauge potential $A^{\mu }$ and
the electric current density $j_{\mu }$. In the $3$-vector notation, we have 
$F^{\mu \nu }F_{\mu \nu }$ $=2(\mathbf{B}^{2}-\mathbf{E}^{2})$ and $F^{\mu
\nu }F_{\mu \nu }^{\star }=4\mathbf{E}\bullet \mathbf{B}$, where $\mathbf{E}$
and $\mathbf{B}$ are the electric and magnetic fields, respectively. The
field equations derived from (\ref{4}) are%
\begin{equation}
\partial ^{\mu }F_{\mu \nu }=4\pi j_{\nu }+\left( \partial ^{\mu }{}\alpha
\right) F_{\mu \nu }^{\star }\text{ \ \ and \ \ }\partial ^{\mu }F_{\mu \nu
}^{\star }=0.  \label{5}
\end{equation}%
The first expression ensures the effect caused due to an axion field.
Alternatively, in \cite{tiwari}, it occurs a generalization of the axion
electrodynamics to contain a local duality symmetry. In this case, it is%
\begin{equation}
\partial ^{\mu }F_{\mu \nu }=4\pi j_{\nu }+\left( \partial ^{\mu }{}\alpha
\right) F_{\mu \nu }^{\star }\text{ \ \ \ \ and \ \ \ \ }\partial ^{\mu
}F_{\mu \nu }^{\star }=4\pi m_{\nu }-\left( \partial ^{\mu }{}\alpha \right)
F_{\mu \nu },  \label{6}
\end{equation}%
where $m_{\mu }$ is a magnetic monopole current density.

In general, the system (\ref{6}) is invariant under the dual transformations 
$\mathfrak{F}\mapsto \mathfrak{TF}$, $\mathfrak{C}\mapsto \mathfrak{TC}$ and 
$\partial _{\mu }\alpha \mapsto \partial _{\mu }\alpha +\partial _{\mu }\phi 
$, where $\mathfrak{F}$, $\mathfrak{C}$ and $\mathfrak{T}$ respectively are
given by%
\begin{equation}
\mathfrak{F}=%
\begin{pmatrix}
F_{\mu \nu } \\ 
F_{\mu \nu }^{\star }%
\end{pmatrix}%
,\text{ \ \ }\mathfrak{C}=%
\begin{pmatrix}
j_{\nu } \\ 
m_{\nu }%
\end{pmatrix}%
\text{ \ \ and \ \ }\mathfrak{T}=%
\begin{pmatrix}
\cos \phi & \sin \phi \\ 
-\sin \phi & \cos \phi%
\end{pmatrix}%
,  \label{7}
\end{equation}%
with $\phi =\phi \left( x^{\mu }\right) $. It is worth reporting that (\ref%
{6}) assumes the pattern (\ref{5}) by putting out the right side of the
second expression, so that $\left( \partial ^{\mu }{}\alpha \right) F_{\mu
\nu }=4\pi m_{\nu }$. We can understand this fact as a solution for the
non-observation of the magnetic monopole in nature. Still by (\ref{6}), it
is clear that the dual axion electrodynamics that satisfies the Maxwell
electromagnetism provides 
\begin{equation}
\left( \partial ^{\mu }{}\alpha \right) F_{\mu \nu }^{\star }=0\text{ \ \ \
\ and \ \ \ \ }\left( \partial ^{\mu }{}\alpha \right) F_{\mu \nu }=4\pi
m_{\nu }.  \label{8}
\end{equation}

On the other hand, the Clifford algebra $\mathcal{C}l(3,1)$ in the $\gamma $%
-formalism is represented by%
\begin{equation}
g_{\mu \nu }=\Upsilon _{\mu }^{AA^{\prime }}\Upsilon _{\nu }^{BB^{\prime
}}\gamma _{AB}\gamma _{A^{\prime }B^{\prime }}.  \label{9}
\end{equation}%
There is an adoption of Einstein's convention and each spinor index runs
from $0$ to $1$ ($0^{\prime }$ to $1^{\prime }$). The metric spinor $\gamma
_{AB}$ and the Infeld-van der Waerden symbols $\Upsilon _{\mu }^{AA^{\prime
}}$ are defined by (\ref{2}) (for a review about the objects, see \cite%
{infeld,cardoso}). We use the metric spinor to lower (or raise) the spinor
indexes: $\xi _{A}=\gamma _{BA}\xi ^{B}$, $\xi ^{A}=\gamma ^{AB}\xi _{B}$.
The object $\gamma _{AB}$ is a skew-symmetric spinor component. In the
matrix form $\left( \gamma _{AB}\right) $, it is%
\begin{equation}
(\gamma _{AB})=%
\begin{pmatrix}
0 & \gamma \\ 
-\gamma & 0%
\end{pmatrix}%
,\text{ \ \ with \ \ }\gamma \doteqdot \left\vert \gamma \right\vert
e^{\Theta i},  \label{10}
\end{equation}%
where $\left\vert \gamma \right\vert $ and $\Theta $ are real-valued
functions of $x^{\mu }$. Spinors and tensors are related by using a
hermitian matrix set $\Upsilon $, such as $v_{\mu }=\Upsilon _{\mu
}^{AA^{\prime }}v_{AA^{\prime }}$ and $v_{AA^{\prime }}=\Upsilon
_{AA^{\prime }}^{\mu }v_{\mu }$. In general, we will denote a complex
conjugation by $(S_{\mu }^{AB^{\prime }})^{\ast }=S_{\mu }^{A^{\prime }B}$,
for any $S_{\mu }^{AB^{\prime }}$.

The covariant derivative of some generic spinors $\xi ^{A}$ and $\zeta _{A}$
are given as it follows%
\begin{equation}
\nabla _{\mu }\xi ^{A}=\partial _{\mu }\xi ^{A}+\Xi _{\mu B}{}^{A}\xi ^{B}%
\text{ \ \ \ \ and \ \ \ \ }\nabla _{\mu }\zeta _{A}=\partial _{\mu }\zeta
_{A}-\Xi _{\mu A}{}^{B}\zeta _{B},  \label{11}
\end{equation}%
with $\Xi _{\mu B}{}^{A}$ being a spinor connection. In the $\gamma $%
-formalism, the covariant derivative $\nabla _{AA^{\prime }}$ is taken by
using the Infeld-van der Waerden symbols: $\nabla _{AA^{\prime }}=\Upsilon
_{AA^{\prime }}^{\mu }\nabla _{\mu }$ and $\nabla ^{AA^{\prime }}=\gamma
^{AB}\gamma ^{A^{\prime }B^{\prime }}\Upsilon _{BB^{\prime }}^{\mu }\nabla
_{\mu }$. In Minkowski universe, $\nabla _{\mu }=\partial _{\mu }$ and $%
\Upsilon _{AA^{\prime }}^{\mu }=\sigma _{AA^{\prime }}^{\mu }/\sqrt{2}$
since $\left\vert \gamma \right\vert =1$. Here $\sigma _{AA^{\prime }}^{\mu
}=$ $(\mathbb{I},\sigma _{AA^{\prime }}^{i})$ where $\sigma _{AA^{\prime
}}^{i}$ are the Pauli matrices and $\mathbb{I}$ the $2\times 2$ unity
matrix. The complex component $\Xi _{\mu A}{}^{A}$ can be written as \cite%
{cardoso}%
\begin{equation}
\Xi _{\mu A}{}^{A}=\partial _{\mu }\ln \left\vert \gamma \right\vert -2i\Xi
_{\mu },\text{ \ \ with \ \ }\Xi _{\mu }\doteqdot -(1/2)\text{Im}\Xi _{\mu
A}{}^{A}.  \label{12}
\end{equation}

In the $\gamma $-formalism, the metric compatibility $\nabla _{\alpha }$ $%
g_{\mu \nu }=0$ leads to the eigenvalue equations \cite{cardoso}%
\begin{equation}
\nabla _{\mu }\gamma _{AB}=i\beta _{\mu }\gamma _{AB}\text{ \ \ \ \ and \ \
\ \ }\nabla _{\mu }\gamma ^{AB}=-i\beta _{\mu }\gamma ^{AB},  \label{13}
\end{equation}%
where $\beta _{\mu }$ is defined by%
\begin{equation}
\beta _{\mu }\doteqdot \partial _{\mu }\Theta +2\Xi _{\mu }.  \label{14}
\end{equation}%
In flat space-time, we can consider the choice $\Xi _{\mu A}{}^{B}=0$. Thus, 
$\Xi _{\mu }=0$ and $\left\vert \gamma \right\vert $ $=const>0$, so that $%
\beta _{\mu }=\partial _{\mu }\Theta $. Here, we will assume $\left\vert
\gamma \right\vert $ $=1$. It is usual the spinors $\xi ^{A}$ and $\zeta
_{A} $ transform under the action of the generalized Weyl group, which has
the following component form:%
\begin{equation}
\Delta _{A}{}^{B}=\sqrt{\rho }e^{i\lambda }\delta _{A}{}^{B},  \label{15}
\end{equation}%
where $\rho >0$ is a real function and $\lambda $ the gauge parameter of the
group.

The $2$-spinor version of the Maxwell tensor and its Hodge dual in the $%
\gamma $-formalism are \cite{cardoso,penrose,carmeli}%
\begin{equation}
2F_{AA^{\prime }BB^{\prime }}=\gamma _{AB}f_{A^{\prime }B^{\prime }}+\gamma
_{A^{\prime }B^{\prime }}f_{AB}\text{ \ \ and \ \ }2F_{AA^{\prime
}BB^{\prime }}^{\star }=i\left( \gamma _{AB}f_{A^{\prime }B^{\prime
}}-\gamma _{A^{\prime }B^{\prime }}f_{AB}\right) ,  \label{16}
\end{equation}%
in which $f_{AB}=f_{\left( AB\right) }$ is named Maxwell spinor. It is
convenient to define the following complex linear combination: $F_{\mu \nu
}^{(\pm )}\doteqdot F_{\mu \nu }\pm iF_{\mu \nu }^{\star }$ so that by
taking the expressions (\ref{16}), we find%
\begin{equation}
F_{AA^{\prime }BB^{\prime }}^{(+)}=\gamma _{A^{\prime }B^{\prime }}f_{AB}%
\text{ \ \ \ \ and \ \ \ \ }F_{AA^{\prime }BB^{\prime }}^{(-)}=\gamma
_{AB}f_{A^{\prime }B^{\prime }}.  \label{17}
\end{equation}%
Let's then take the complex version of Maxwell's equations, i. e.,%
\begin{equation}
\nabla ^{\mu }F_{\mu \nu }^{(\pm )}=4\pi j_{\nu }.  \label{18}
\end{equation}%
If we consider (\ref{17}) as well as the eigenvalue equations (\ref{13}),
the complex form (\ref{18}) yields%
\begin{equation}
\nabla _{A^{\prime }}^{B}f_{AB}=2\pi j_{AA^{\prime }}+i\beta _{A^{\prime
}}^{B}f_{AB}.  \label{19}
\end{equation}%
In the $\varepsilon $-formalism, as the equation (\ref{18}) with magnetic
source term $\nabla ^{\mu }F_{\mu \nu }^{(\pm )}=4\pi \left( j_{\nu }\pm
im_{\nu }\right) $ provides $\nabla _{A^{\prime }}^{B}f_{AB}=2\pi \left(
j_{AA^{\prime }}+im_{AA^{\prime }}\right) $, it follows the definition
elaborated in \cite{amkafs}, namely%
\begin{equation}
\beta _{A^{\prime }}^{B}f_{AB}\doteqdot 2\pi m_{AA^{\prime }}.  \label{20}
\end{equation}%
In world representation, the definition (\ref{20}) implies%
\begin{equation}
\beta ^{\mu }F_{\mu \nu }^{(\pm )}=4\pi m_{\nu }.  \label{21}
\end{equation}%
The expression (\ref{21}) is the master equation to identify the space-time
phase with the axion field.

In Minkowski space-time, the $\beta $-vector is $\beta _{\mu }=\partial
_{\mu }\Theta $ so that the equation (\ref{21}) becomes $\left( \partial
^{\mu }\Theta \right) F_{\mu \nu }^{(\pm )}=4\pi m_{\nu }$. Now, if we
specify $F_{\mu \nu }^{(\pm )}$ at terms of the Maxwell tensor and its Hodge
dual, we obtain the following pair of equations%
\begin{equation}
\left( \partial ^{\mu }\Theta \right) F_{\mu \nu }^{\star }=0\text{ \ \ and
\ \ }\left( \partial ^{\mu }\Theta \right) F_{\mu \nu }=4\pi m_{\nu }.
\label{22}
\end{equation}%
It is clear that the expressions (\ref{22}) and (\ref{8}) are formally
identical so that $\Theta $ $\sim $ $\alpha $ establishes in \cite{amkafs}.

On the Lagrangian viewpoint in flat universe, the Maxwell Lagrangian in the $%
\gamma $-formalism is given by%
\begin{equation}
\mathcal{L}_{M}=2F^{\mu \nu }F_{\mu \nu }=\text{Re}\left[ \gamma ^{AC}\gamma
^{BD}f_{AB}f_{CD}\right] .  \label{23}
\end{equation}%
If we explicit the $\gamma $-terms, the right side of (\ref{23}) leads to%
\begin{equation}
\text{Re}\left[ \varepsilon ^{AC}\varepsilon ^{BD}f_{AB}f_{CD}\right] \cos
\left( 2\Theta \right) +\text{Im}\left[ \varepsilon ^{AC}\varepsilon
^{BD}f_{AB}f_{CD}\right] \sin \left( 2\Theta \right) .  \label{24}
\end{equation}%
By taking the approximation $\Theta \simeq 0$ ($\gamma ^{AB}$ $\simeq $ $%
\varepsilon ^{AB}$), we have%
\begin{equation}
\text{Re}\left[ \varepsilon ^{AC}\varepsilon ^{BD}f_{AB}f_{CD}\right] \simeq
2F^{\mu \nu }F_{\mu \nu }\text{ \ \ and \ \ Im}\left[ \varepsilon
^{AC}\varepsilon ^{BD}f_{AB}f_{CD}\right] \simeq 2F^{\mu \nu }F_{\mu \nu
}^{\star },  \label{25}
\end{equation}%
whence, due to (\ref{24}), we find of (\ref{23}) an effective Chern-Simons
theory:%
\begin{equation}
\mathcal{L}_{M}\simeq 2F^{\mu \nu }F_{\mu \nu }+4\Theta F^{\mu \nu }F_{\mu
\nu }^{\star }.  \label{26}
\end{equation}%
Hence, when $\Theta \simeq 0$ the invariant form Re$\left[ f^{AB}f_{AB}%
\right] $ in the $\gamma $-formalism generates an axion-like space-time
phase-photon coupling.

\section{Axion-like Phase-Fermion Coupling and Peccei-Quinn Pseudoparticle
Behavior}

Similarly, as we have seen for Maxwell's theory, in this section we will
study the Dirac's theory to show that the interactions between the phase and
some fermions are axion-like. Afterwards, we will generalize the Weyl
transformation to a general spin transformation that carries Peccei-Quinn
and Weyl spin rotations simultaneously. With the Weyl-PQ transformation
established, we will demonstrate that the space-time phase behaves as a
pseudoscalar (such as an axion-like pseudoparticle) under the action of PQ
group.

\subsection{Axion-like Phase-fermion Coupling}

For a comprehensive review about Dirac's theory in the Infeld-van der
Waerden's formalisms, see \cite{infeld,bade,dcardoso}. Let's follow the Ref. 
\cite{penrose} to present the first pair of Dirac equations. In the 2-
spinor formalism, the fermion dynamics, in a generic space-time, is
described by the pair%
\begin{equation}
\nabla ^{AA^{\prime }}\psi _{A}=\mu \chi ^{A^{\prime }}\text{ \ \ and\ \ \ }%
\nabla _{AA^{\prime }}\chi ^{A^{\prime }}=-\mu \psi _{A},  \label{27}
\end{equation}%
where $\psi _{A}$ and $\chi ^{A^{\prime }}$ are Weyl $2$-spinors with
opposite chiralities, and $\mu \doteqdot m/\sqrt{2}$ the normalized rest
mass.

In the $\gamma $-formalism, because we have the eigenvalue equations (\ref%
{13}), the duo (\ref{27}) is equivalent to%
\begin{equation}
\nabla _{AA^{\prime }}\psi ^{A}+i\beta _{AA^{\prime }}\psi ^{A}=-\mu \chi
_{A^{\prime }}\text{ \ \ and \ \ }\nabla ^{AA^{\prime }}\chi _{A^{\prime
}}+i\beta ^{AA^{\prime }}\chi _{A^{\prime }}=\mu \psi ^{A}.  \label{28}
\end{equation}%
The Dirac fields which satisfy (\ref{27}) and (\ref{28}) respectively (as
well as their complex conjugates) are $\mathfrak{D}=\left\{ \left( \psi
_{A},\chi ^{A^{\prime }}\right) ,\left( \chi ^{A},\psi _{A^{\prime }}\right)
\right\} $ and $\mathfrak{D}^{\Updownarrow }=\left\{ \left( \psi ^{A},\chi
_{A^{\prime }}\right) ,\left( \chi _{A},\psi ^{A^{\prime }}\right) \right\} $%
. By using the metric spinors, depending on the formalism considered, $%
\mathfrak{D}^{\Updownarrow }$ is obtained from $\mathfrak{D}$. In the $%
\varepsilon $-formalism, when putting out the $\beta $-terms, we obtain the
equivalent equations of (\ref{28}).

If we also consider (\ref{14}) and (\ref{28}) as the reference \cite{adshead}%
, where the axion-fermion coupling has been explicitly written in the index $%
2$-spinor formalism, we clearly see an axion-like coupling for the system $%
\mathfrak{D}^{\Updownarrow }$. In fact, the $\beta $-terms of (\ref{28}) are
decomposed as%
\begin{equation}
\beta _{AA^{\prime }}\psi ^{A}=\underset{\text{axion-like}}{\underbrace{\psi
^{A}\partial _{AA^{\prime }}\Theta }}+\underset{\text{background}}{%
\underbrace{2\Xi _{AA^{\prime }}\psi ^{A}}}\text{ \ \ and \ \ }\beta
^{AA^{\prime }}\chi _{A^{\prime }}=\underset{\text{axion-like}}{\underbrace{%
\chi _{A^{\prime }}\partial ^{AA^{\prime }}\Theta }}+\underset{\text{%
background}}{\underbrace{2\Xi ^{AA^{\prime }}\chi _{A^{\prime }}}},
\label{29}
\end{equation}%
where we used (\ref{14}) to decompose the $\beta $-terms and \cite{adshead}
to identify the axion-like couplings.

It is important mentioning that solely $\mathfrak{D}^{\Updownarrow }$
couples in an explicit way with $\Theta $, while $\mathfrak{D}$ does not.
Usually, the $4$-component Dirac spinor is defined by taking $\mathfrak{D}%
_{1}$ $=$ $\left( \psi _{A},\chi ^{A^{\prime }}\right) $, since $\mathfrak{D}%
_{1}\mapsto $ $e^{i\lambda }\mathfrak{D}_{1}$ under the action of the Weyl
group $\Delta _{A}^{\circ }{}^{B}$ (where $\rho $ $=1$ in (\ref{15})). On
the other hand, $\mathfrak{D}_{1}^{\Updownarrow }$ $=\left( \psi ^{A},\chi
_{A^{\prime }}\right) $ transforms as $\mathfrak{D}_{1}^{\Updownarrow
}\mapsto $ $e^{-i\lambda }\mathfrak{D}_{1}^{\Updownarrow }$. We can invert
this election for a sign redefinition $\lambda $ $\rightarrow $ $-\lambda $
in (\ref{15}).

On the Lagrangian viewpoint, the axion-like coupling is seen through the
manipulation%
\begin{equation}
\psi _{A^{\prime }}\nabla ^{AA^{\prime }}\psi _{A}+\chi ^{A}\nabla
_{AA^{\prime }}\chi ^{A^{\prime }}=\psi ^{A^{\prime }}\nabla _{A^{\prime
}A}\psi ^{A}+\chi _{A}\nabla ^{AA^{\prime }}\chi _{A^{\prime }}+i\psi
^{A}\psi ^{A^{\prime }}\beta _{AA^{\prime }}+i\chi _{A^{\prime }}\chi
_{A}\beta ^{AA^{\prime }}\subset \mathcal{L}_{D},  \label{30}
\end{equation}%
where $\mathcal{L}_{D}$ is the Dirac Lagrangian. It follows then by (\ref{14}%
)%
\begin{equation}
\psi ^{A}\psi ^{A^{\prime }}\beta _{AA^{\prime }}+\chi _{A^{\prime }}\chi
_{A}\beta ^{AA^{\prime }}=\underset{\text{axion-like couplings}}{\underbrace{%
\psi ^{A}\psi ^{A^{\prime }}\partial _{AA^{\prime }}\Theta +\chi _{A^{\prime
}}\chi _{A}\partial ^{AA^{\prime }}\Theta }}+\underset{\text{background
couplings}}{\underbrace{2\left( \psi ^{A}\psi ^{A^{\prime }}\Xi _{AA^{\prime
}}+\chi _{A^{\prime }}\chi _{A}\Xi ^{AA^{\prime }}\right) }}.  \label{31}
\end{equation}%
The decomposition (\ref{31}) naturally provides an axion-like interaction
term in Dirac's theory.

\subsection{Weyl-PQ Spin Transformations: Space-time Phase as a Peccei-Quinn
Pseudoparticle}

The space-time phase has been identified here with the axion field based on
its coupling with Dirac field, as with the Maxwell field in \cite{amkafs}.
Individually, those identifications are not enough, since the behavior of $%
\Theta $ under chiral rotation has not been established. As we have pointed
out in the previous subsection, the Weyl $2$-spinors, according with the
action of the Weyl group, defines the Dirac spinor. However, we have the
chiral rotations too, which rotate the Dirac spinors in the following manner:%
\begin{equation}
\Psi \text{ }\widetilde{\mapsto }\text{ }\widetilde{\Psi }=e^{i\zeta \gamma
_{5}}\Psi .  \label{32}
\end{equation}%
In here, we will denote respectively by $n$ $\widehat{\mapsto }$ $\widehat{n}
$ and $n$ $\widetilde{\mapsto }$ $\widetilde{n}$, the Weyl and \textrm{PQ}
transformations. The original Peccei-Quinn procedure \cite{peccei} states
that, under a chiral rotation, an axion-like pseudoscalar $\theta $
transforms as%
\begin{equation}
\theta \text{ }\widetilde{\mapsto }\text{ }\theta -2\zeta .  \label{33}
\end{equation}%
The transformation (\ref{33}) solves the \textrm{CP}-problem. Our strategy
will be to identify $\Theta $ with $\alpha $, using a similar idea to the
one developed by Infeld and van der Waerden.

Long ago, Infeld and van der Waerden considered the gauge behavior of the
electromagnetic potential $A_{\mu }$, i. e.,%
\begin{equation}
A_{\mu }\text{ }\widehat{\mapsto }\text{ }A_{\mu }-\partial _{\mu }\lambda .
\label{34}
\end{equation}%
Once they knew that the object $\Xi _{\mu }$ transforms as $\Xi _{\mu }$ $%
\widehat{\mapsto }$ $\Xi _{\mu }-\left( 1/2\right) $Im$\left[ \partial _{\mu
}\ln \left( \det \Delta _{A}^{\circ }{}^{B}\right) \right] $ (see for
example the Ref. \cite{cardoso}), they would put the determinant $\det
\Delta _{A}^{\circ }{}^{B}=e^{2\lambda i}$ of the Weyl group and so to find%
\begin{equation}
\Xi _{\mu }\text{ }\widehat{\mapsto }\text{ }\Xi _{\mu }-\partial _{\mu
}\lambda .  \label{35}
\end{equation}%
Because (\ref{34}) look likes (\ref{35}), Infeld and van der Waerden
suggested the identification $\Xi _{\mu }$ $\sim $ $A_{\mu }$.

Our strategy is to verify if $\Theta $ behaves as (\ref{33}) when
considering a \textrm{PQ} transformation. A similar problem was solved in 
\cite{toyoda}, where the author worked $\Xi _{\mu }$ as being a mixture of
polar and axial vectors. Our first step is to find spin transformations that
represent simultaneously Weyl and \textrm{PQ} rotations. In the $2$-spinor
language, the translation for chiral rotation is $\widetilde{\psi }%
_{A}=e^{i\zeta }\psi _{A}$ and $\widetilde{\chi }^{A^{\prime }}=e^{-i\zeta
}\chi ^{A^{\prime }}$. Then we need to obtain the spin transformations
where, in general, $\overline{\psi }_{A}=e^{i\left( \lambda +\zeta \right) }$%
\ and $\overline{\chi }^{A^{\prime }}=e^{i\left( \lambda -\zeta \right)
}\chi ^{A^{\prime }}$. The notation ($\overline{n}$) denotes the composition
of the Weyl ($\widehat{n}$)-PQ ($\widetilde{n}$) transformations.

Let's consider the general spin transformation of the metric spinor:%
\begin{equation}
\overline{\gamma }_{AB}=\Delta _{A}{}^{C}\Delta _{B}{}^{D}\gamma _{CD}.
\label{36}
\end{equation}%
Now, we decompose (\ref{36}) as it follows%
\begin{equation}
\Delta _{A}{}^{B}=\Delta _{A}^{\circ }{}^{C}\Delta _{C}^{\bullet }{}^{B}=%
\sqrt{\Delta ^{\circ }\Delta ^{\bullet }}\delta _{A}{}^{B},  \label{37}
\end{equation}%
where $\Delta ^{\circ }$ and $\Delta ^{\bullet }$ are the determinants of
the Weyl $(\Delta _{A}^{\circ }{}^{B})$ and \textrm{PQ} $(\Delta
_{A}^{\bullet }{}^{B})$ rotations. Under a Weyl transformation, we have $%
\widehat{\gamma }_{AB}=\Delta ^{\circ }\gamma _{AB}$ while under a \textrm{PQ%
} transformation, we supose%
\begin{equation}
\widetilde{\gamma }_{AB}=\Delta ^{\bullet }(\widetilde{\gamma }\varepsilon
_{AB}).  \label{38}
\end{equation}%
Through the previous considerations, the metric spinor transformation (\ref%
{36}) becomes%
\begin{equation}
\overline{\gamma }_{AB}=\Delta ^{\circ }\Delta ^{\bullet }(\widetilde{\gamma 
}\varepsilon _{AB}).  \label{39}
\end{equation}%
As we will verify later, a \textrm{PQ} transformation can be implemented by
demanding%
\begin{equation}
\overline{\gamma }_{AB}=\Delta ^{\circ }\gamma _{AB}\text{ \ \ }%
\Leftrightarrow \text{ \ \ }\widetilde{\gamma }=(\Delta ^{\bullet
})^{-1}\gamma .  \label{40}
\end{equation}%
From (\ref{40}), it is clear $\overline{\gamma }_{AB}=\widehat{\gamma }_{AB}$%
. By implementing $\overline{\gamma }^{AB}\overline{\gamma }_{AB}=\gamma
^{AB}\gamma _{AB}$, we deduce the expressions%
\begin{equation}
\overline{\gamma }^{AB}=\left( \Delta ^{\circ }\right) ^{-1}\gamma ^{AB}%
\text{ \ \ and \ \ }\widetilde{\gamma }^{-1}=\Delta ^{\bullet }\gamma ^{-1}.
\label{41}
\end{equation}

The transformation of generic spinors $v_{A}$ and $u^{A}=\gamma ^{AB}u_{B}$
are then given by%
\begin{equation}
\overline{v}_{A}=\sqrt{\Delta ^{\circ }\Delta ^{\bullet }}v_{A}\text{ \ \
and \ \ }\overline{u}^{A}=\sqrt{\left( \Delta ^{\circ }\right) ^{-1}\Delta
^{\bullet }}u^{A}.  \label{42}
\end{equation}%
Since $\Delta ^{\circ }=e^{2\lambda i}$ and $\Delta ^{\bullet }=e^{2\zeta i}$%
, the $2$-component fermions transform as%
\begin{equation}
\overline{\psi }_{A}=e^{i\left( \lambda +\zeta \right) }\psi _{A}\text{ \ \
and \ \ }\overline{\chi }^{A^{\prime }}=e^{i\left( \lambda -\zeta \right)
}\chi ^{A^{\prime }},  \label{43}
\end{equation}%
or particularly $\widehat{\psi }_{A}=e^{i\lambda }\psi _{A}$ ($\widehat{\chi 
}^{A^{\prime }}=e^{i\lambda }\chi ^{A^{\prime }}$) (Weyl) and $\widetilde{%
\psi }_{A}=e^{i\zeta }\psi _{A}$ ($\widetilde{\chi }^{A^{\prime
}}=e^{-i\zeta }\chi ^{A^{\prime }}$) (\textrm{PQ}).

As we have observed, the Weyl-\textrm{PQ} transformation is possible if it
satisfies $\widetilde{\gamma }=(1/\Delta ^{\bullet })\gamma $. In Minkowski
space-time, we have $\gamma =e^{i\Theta }$, such that under a \textrm{PQ}
rotation, we obtain $e^{i\widetilde{\Theta }}=(e^{i\Theta }/\Delta ^{\bullet
})$. Therefore $\widetilde{\Theta }=\Theta +i\ln \Delta ^{\bullet }+2n\pi $,
so that%
\begin{equation}
\Theta \text{ }\widetilde{\mapsto }\text{ }\underset{\text{PQ-behavior}}{%
\underbrace{\Theta -2\zeta }}+2n\pi ,\text{ \ \ }n\in \mathbb{Z},  \label{44}
\end{equation}%
since $\Delta ^{\bullet }=e^{2\zeta i}$, with $\mathbb{Z}$ being the set of
integers. We have demonstrated that $\Theta $ satisfies the \textrm{PQ}
requirement (\ref{33}). We must note that the invariant form $\overline{u}%
^{A}\overline{v}_{A}$ is broken in our formulation: $\overline{u}^{A}%
\overline{v}_{A}=e^{2\zeta i}u^{A}v_{A}$. This property is a formal
declaration of no chiral symmetry for massive fermion terms.

\section{Maxwell-Dirac System: Magnetic Monopole Current and its Dyonic
Charge}

When coupled with Maxwell fields, the Dirac equations in curved space-times
are taken when substituting $\nabla _{AA^{\prime }}$ by $\nabla _{AA^{\prime
}}-ieA_{AA^{\prime }}$ in (\ref{28}), with $A_{\mu }$ being an
electromagnetic potential component. It follows the expression (\ref{19})
rewritten as%
\begin{equation}
\left[ \nabla _{A^{\prime }}^{B}-i\left( eA_{A^{\prime }}^{B}+\partial
_{A^{\prime }}^{B}\Theta +2\Xi _{A^{\prime }}^{B}\right) \right]
f_{AB}=e\left( \psi _{A}\psi _{A^{\prime }}+\chi _{A}\chi _{A^{\prime
}}\right) ,  \label{45}
\end{equation}%
where we used (\ref{14}) and the electric current density \cite{penrose}%
\begin{equation}
j_{AA^{\prime }}=q\left( \psi _{A}\psi _{A^{\prime }}+\chi _{A}\chi
_{A^{\prime }}\right) ,\text{ \ \ with \ \ }q\doteqdot \frac{e}{2\pi },
\label{46}
\end{equation}%
being $e$ the electric charge.

Let's consider the equations (\ref{22}) in curved space-time:%
\begin{equation}
\nabla ^{\mu }F_{\mu \nu }^{(\pm )}=4\pi j_{\nu }\text{ \ \ and \ \ }\beta
^{\mu }F_{\mu \nu }^{(\pm )}=4\pi m_{\nu }.  \label{47}
\end{equation}%
If we apply the covariant derivative on the last one of (\ref{47}), we will
obtain%
\begin{equation}
W^{\mu \nu }F_{\mu \nu }+8\pi \left( \beta ^{\mu }j_{\mu }+\nabla ^{\mu
}m_{\mu }\right) =0\text{ \ \ \ \ and \ \ \ }W^{\mu \nu }F_{\mu \nu }^{\star
}=0,  \label{48}
\end{equation}%
where $W_{\mu \nu }\doteqdot 2\partial _{\lbrack \mu }\Xi _{\nu ]}$ is an
Infeld-van der Waerden curvature bivector. Still, $\beta ^{\lbrack \mu
}\beta ^{\nu ]}=0$ implies $\beta ^{\mu }m_{\mu }=0$.

In flat space-time, when remembering the definition $\Xi _{\mu }=-(1/2)\func{%
Im}\Xi _{\mu A}{}^{A}$, the first equation of (\ref{48}) leads to $\partial
^{\mu }m_{\mu }=-j_{\mu }\partial ^{\mu }\Theta $. If we require the null
divergence for electric sources, namely $\partial ^{\mu }j_{\mu }=0$, it
follows%
\begin{equation}
m_{\mu }=-\left( \Theta -C\right) j_{\mu },  \label{49}
\end{equation}%
where $C$ is a constant. Once $\beta ^{\mu }m_{\mu }=0$ and, in general, $%
\Theta \left( x^{\mu }\right) \neq C$, in Minkowski universe $-\left( \Theta
-C\right) j_{\mu }\partial ^{\mu }\Theta =0$, which implies $j_{\mu
}\partial ^{\mu }\Theta =0$ so that $\partial ^{\mu }m_{\mu }=0$. Therefore,
in flat background, we have the divergence equations%
\begin{equation}
\partial ^{\mu }j_{\mu }=0=\partial ^{\mu }m_{\mu },  \label{50}
\end{equation}%
as well as the orthogonality relationships%
\begin{equation}
j_{\mu }\partial ^{\mu }\Theta =0=m_{\mu }\partial ^{\mu }\Theta .
\label{51}
\end{equation}

If we take (\ref{46}), the $2$-spinor form of (\ref{49}) yields%
\begin{equation}
m_{AA^{\prime }}=g\left( \psi _{A}\psi _{A^{\prime }}+\chi _{A}\chi
_{A^{\prime }}\right) ,\text{ \ \ with \ \ }g\doteqdot -\frac{e}{2\pi }%
\left( \Theta -C\right) .  \label{52}
\end{equation}%
The term $g$ in (\ref{52}) can be understood as an effective charge and
interpreted as a dyonic charge \cite{zwanziger,schwinger}. Under a PQ
transformation, our dyon charge behaves as%
\begin{equation}
g\text{ }\widetilde{\mapsto }\text{ }-\frac{e}{2\pi }\left( \Theta -2\zeta
+2n\pi -C\right) .  \label{53}
\end{equation}%
When $\zeta =$ $n\pi $ with $n$ $\in \mathbb{Z}$, we have $\widetilde{g}=g$.
Another case, where $\widetilde{g}=g$, it is when $2\zeta $ $=C=2n\pi $. In
the last one, we derive a quantized dyon charge given by%
\begin{equation}
g=-e\left( \frac{\Theta }{2\pi }-n\right) .  \label{54}
\end{equation}%
The expression (\ref{54}) is identical to the found in the named Witten
effect \cite{witten}.

From usual axion electrodynamics, the Witten effect is obtained when
demanding $\mathbf{\nabla }\bullet \mathbf{B}\neq 0$. However, we must note
that our charge (\ref{54}) is not associated with the monopole equation. To
be specific, since $\left( \partial ^{\mu }\Theta \right) F_{\mu \nu }=4\pi
m_{\nu }$ and $\partial ^{\mu }F_{\mu \nu }^{\star }=0$, the monopole
density is proportional to $\left( \mathbf{\nabla }\Theta \right) \bullet 
\mathbf{E}$ while $\mathbf{\nabla }\bullet \mathbf{B}=0$. Hence, the global
definition $\mathbf{B}=\mathbf{\nabla }\times \mathbf{A}$ for magnetic field
remains valid.

\section{Concluding Remarks and Outlook}

By means of \cite{amkafs}, we recapitulated that Maxwell's theory generates
an axion electrodynamics when defining magnetic monopole currents in a
suitable way. Especially in this article, we propose to promote the idea
which conjectures the axion as a physical manifestation of a space-time
internal freedom, namely a phase in the metric spinor. In view of outhers
interactions of the axion field, we obtained axion-like space-time
phase-fermion couplings in the usual Dirac's theory. Thus, we have shown
through Maxwell (Ref. \cite{amkafs}) and Dirac theories that Infeld-van der
Waerden's $\gamma $-formalism provides an axionic classical sector, where a
metric spinor phase played the axion role.

However, under \textrm{PQ} rotations, the behavior of this phase was not
established. So, we elaborated a Weyl-\textrm{PQ} approach in which the 
\textrm{PQ} rotations are implemented in the usual spin transformations,
concluding that the phase behaves geometrically as an axion pseudoparticle.
On the final part of this investigation, we studied the Dirac-Maxwell system
to obtain a $2$-spinor form for the magnetic monopole current. The
proposition showed that the monopole has a dyon charge and acquires a Witten
effect pattern in flat space-time when considering a chiral symmetry. In
future investigations, the researchers must consider a rigorous formulation
of the Weyl-\textrm{PQ} transformations in Infeld-van der Waerden's
formalisms, as well as a more accurate analysis of the Dirac-Maxwell system
and its dyon structure.

In accordance with theoretical scope, it is indispensable to research some
experimental and phenomenological implications as, for example, we suggest
in what follows:

\textit{Magnetic (dyon) charges bounds in local dual axion electrodynamics.-}
Recently, stronger bounds on magnetic and dyon charges have been analyzed in
the Refs. \cite{bojowald,addazi} using non-associative quantum mechanics. As
can be seen in \cite{addazi}, when considering dyonic charge for protons or
proton-electrons, the authors found the bounds $g\leq 7,38\times
10^{-36}g_{Dirac}$ (protons) and $g\leq 1,5\times 10^{-8}g_{Dirac}$
(proton-electrons) by means of the average magnetic field of the Moon, being 
$g_{Dirac}$ the Dirac magnetic charge. The results obtained in \cite{addazi}
(as also in \cite{bojowald}) are derived using the Gauss law for $\mathbf{%
\nabla }\bullet \mathbf{B\neq 0}$, namely $\mathcal{G}\sim \int \left[ 
\mathbf{\nabla }\bullet \mathbf{B}\left( r\right) \right] r^{2}dr$, when
assuming a static magnetic field and $\mathcal{G}$ spherically symmetric.
However, the local dual axion electrodynamics claims that the magnetic
monopole density is absent, in the sense of $\mathbf{\nabla }\bullet \mathbf{%
B=0}$, because $\left( \mathbf{\nabla }\Theta \right) \bullet \mathbf{E}$
eliminates the monopole term. Therefore the correct formula to use would be%
\begin{equation}
\mathcal{G}\sim \int \int \int \left[ \left( \mathbf{\nabla }\Theta \right)
\bullet \mathbf{E}\right] dV.  \label{mon}
\end{equation}%
The expression (\ref{mon}) may be used posterioly to find the bounds of
magnetic (dyon) charge in local dual axion electrodynamics similarly as done
in the Refs. \cite{bojowald,addazi}.

\textit{Axion-photon corversion.-} The conversion of the photons into axions
is obtained when considering the Euler-Heisenberg's theory of quantum
electrodynamics \cite{raffelt2}. The theory which describes this process
takes into account the Lagrangian $\mathcal{L}_{M}+\mathcal{L}_{CS}+\mathcal{%
L}_{EH}$, where the term $\mathcal{L}_{EH}\sim \left( F^{\mu \nu }F_{\mu \nu
}\right) ^{2}+(7/4)\left( F^{\mu \nu }F_{\mu \nu }^{\star }\right) ^{2}$
represents the Euler-Heisenberg's theory in the weak field regime. In the
Ref. \cite{amkafs}, we considered the Maxwell's theory in the $\gamma $%
-formalism to show that, under the approximation of $\Theta $\ very small,
it effectively behaves as a Maxwell-Chern-Simons theory. It is necessary
note that our results only provided formal correspondences between the axion
and the phase without presenting any substantial modification on the
phenomenological viewpoint.

We believe that to verify some modification in current or future physics, we
could consider the effective Euler-Heisenberg's theory in the $\gamma $%
-formalism to observe a possible correction in quantum sector since the term 
$\mathcal{L}_{M}+\mathcal{L}_{EH}$ would approximate to $\mathcal{L}_{M}+%
\mathcal{L}_{CS}+\mathcal{L}_{EH}+\Delta \mathcal{L}_{EH}$, with $\Delta 
\mathcal{L}_{EH}$ the correction in quantum level. Private calculations
indicate%
\begin{eqnarray}
\mathcal{L}_{EH} &\sim &\left( \text{Re}\Sigma \right) ^{2}\left[ \cos
^{2}\left( 2\Theta \right) +\frac{7}{4}\sin ^{2}\left( 2\Theta \right) %
\right] +\left( \text{Im}\Sigma \right) ^{2}\left[ \sin ^{2}\left( 2\Theta
\right) +\frac{7}{4}\cos ^{2}\left( 2\Theta \right) \right]   \notag \\
&&-\frac{3}{2}\text{Re}\Sigma \text{Im}\Sigma \cos \left( 2\Theta \right)
\sin \left( 2\Theta \right) ,  \label{aux}
\end{eqnarray}%
where $\Sigma =\varepsilon ^{AC}\varepsilon ^{BD}f_{AB}f_{CD}$. Then,
similarly as we proceed to derive (\ref{26}), if we consider the
aproximation $\Theta \simeq 0$, we obtain $\Delta \mathcal{L}_{EH}\propto
\Theta F^{\mu \nu }F_{\mu \nu }F^{\alpha \beta }F_{\alpha \beta }^{\star }$
so that, perhaps, we can confront the implications of this term with, for
example, astrophysical phenomena (for instance, see \cite{dessert} for a
recent work).

\textbf{Data Availability Statement:} Data sharing not applicable to this
article as no datasets were generated or analysed during the current study.

\end{document}